\documentclass[aps,prb,twocolumn,showpacs]{revtex4}

\usepackage{graphicx}
\usepackage{amsmath}
\usepackage{amssymb}

\begin{document}

\title{Pseudogap-generated a coexistence of Fermi arcs and Fermi pockets in cuprate superconductors}

\author{Huaisong Zhao}

\affiliation{College of Physics, Qingdao University, Qingdao 266071, China}

\author{Deheng Gao and Shiping Feng\footnote{Corresponding author. E-mail: spfeng@bnu.edu.cn}}


\affiliation{Department of Physics, Beijing Normal University, Beijing 100875, China}

\begin{abstract}
One of the most intriguing puzzle is why there is a coexistence of Fermi arcs and Fermi pockets in the pseudogap phase of cuprate superconductors? This puzzle is calling for an explanation. Based on the $t$-$J$ model in the fermion-spin representation, the coexistence of the Fermi arcs and Fermi pockets in cuprate superconductors is studied by taking into account the pseudogap effect. It is shown that the pseudogap induces an energy band splitting, and then the poles of the electron Green's function at zero energy form two contours in momentum space, however, the electron spectral weight on these two contours around the antinodal region is gapped out by the pseudogap, leaving behind the low-energy electron spectral weight only located at the disconnected segments around the nodal region. In particular, the tips of these disconnected segments converge on the hot spots to form the closed Fermi pockets, generating a coexistence of the Fermi arcs and Fermi pockets. Moreover, the single-particle coherent weight is directly related to the pseudogap, and grows linearly with doping. The calculated result of the overall dispersion of the electron excitations is in qualitative agreement with the experimental data. The theory also predicts that the pseudogap-induced peak-dip-hump structure in the electron spectrum is absent from the hot-spot directions.
\end{abstract}


\maketitle

\section{Introduction}

Superconductivity in cuprate superconductors is realized when charge carriers are doped into a parent Mott insulating state \cite{Bednorz86}. This Mott insulating state emerges to be due to the strong electron correlation \cite{Anderson87,Phillips10}, leading to that an energy gap called the pseudogap exists \cite{Hufner08,Timusk99} above the superconducting (SC) transition temperature $T_{\rm c}$ but below the pseudogap crossover temperature $T^{*}$. In this case, it is widely believed that understanding the pseudogap regime in cuprate superconductors is thought to be key to understanding the high-$T_{\rm c}$ phenomenon in general. An important component of that understanding will be the determination of the particular characteristics of the low-energy electron excitations in the normal-state that evolve into the SC-state for the temperatures $T<T_{\rm c}$ \cite{Hufner08,Timusk99,Damascelli03,Campuzano04,Carbotte11}. It is therefore critically important to know the exact nature of the electron Fermi surface (EFS).

Experimentally, the early angle-resolved photoemission spectroscopy (ARPES) experimental data showed that although the normal-state of cuprate superconductors in the pseudogap phase is metallic, the electron spectral weight around the antinodal region of the Brillouin zone (BZ) is suppressed, and then EFS is truncated to four disconnected Fermi arcs centered on the nodal region  \cite{Marshall96,Norman98,Kanigel06,Yoshida06,Kanigel07,Shi08,Nakayama09,Yoshida09,Sassa11,Meng11,Reber12}. However, the recent improvements in the resolution of the ARPES experiments allowed to resolve additional features in the ARPES spectrum. Among these new achievements is the observation of the Fermi pockets in the pseudogap phase of cuprate superconductors \cite{Yang08,Chang08,Meng09,Yang11}. In particular, the ARPES experiments indicated that the Fermi pockets appear to coexist with the Fermi arcs, and then the area of the Fermi pockets is strongly dependent on the doping concentration \cite{Meng09}. On the other hand, the Fermi pockets have been also observed by the quantum oscillation measurements \cite{Nicolas07,LeBoeuf07,Sebastian08,Chan16}. The combined these ARPES and quantum oscillation experimental data thus provide dramatic new insights into the pseudogap phase and elucidate how the electron excitations differ for different values of the electron momentum and the doping concentration \cite{Marshall96,Norman98,Kanigel06,Yoshida06,Kanigel07,Shi08,Nakayama09,Yoshida09,Sassa11,Meng11,Reber12,Yang08,Chang08,Meng09,Yang11,Nicolas07,LeBoeuf07,Sebastian08,Chan16}.

To date, the origin of the coexistence of the Fermi arcs and Fermi pockets in cuprate superconductors is still debated. In one class of the theories, it has been indicated that a finite pseudogap acts to deform a continuous EFS contour in momentum space to form a coexistence of the Fermi arcs and Fermi pockets \cite{Yang06,Valenzuela07}. Moreover, a hybridization phenomenology has been suggested to describe the pseudogap state \cite{LeBlanc14}, where a momentum independent pseudogap opens along the bonding dispersion which results in symmetric bands relative to the energy of the antibonding dispersion, leading to the appearance of two contours in momentum space to form the Fermi pockets. On the other side is a class of the theories, where the origin of the coexistence of the Fermi arcs and Fermi pockets is thought to be a doping dependent EFS reconstruction due to charge order \cite{Chakravarty03}. In particular, it has been argued that the strong electron correlations in the system can also produce the Fermi pockets \cite{Granath10}. However, one obvious discrepancy is that the spectral weight on the back side of the Fermi pocket near the nodal region is zero from these phenomenological theories, which is at odds with the ARPES experimental results \cite{Meng09}. In our recent work \cite{Feng16}, we have studied the nature of charge order and its evolution with doping in the normal-state pseudogap phase of cuprate superconductors based on the $t$-$J$ model in the fermion-spin representation, and shown that the charge-order state \cite{Comin15,Campi15,Comin14,Wu11,Chang12,Ghiringhelli12,Neto14,Comin15a,Hashimoto15} is driven by the pseudogap-induced Fermi-arc instability, with a characteristic wave vector corresponding to the hot spots on the Fermi arcs rather than the antinodal nesting vector. Furthermore, we have shown that the Fermi arc, charge order, and pseudogap in cuprate superconductors are intimately related each other, and all of them emanates from the electron self-energy due to the interaction between electrons by the exchange of spin excitations. In this paper, we try to discuss the origin of the coexistence of the Fermi arcs and Fermi pockets in the pseudogap phase of cuprate superconductors along with this line. Our results show that in the pseudogap phase, the pseudogap induces an energy band splitting, and then the poles of the electron Green's function at zero energy form two continuous contours in momentum space, however, the electron spectral weight on these two continuous contours around the antinodal region is gapped out by the momentum dependence of the pseudogap, and then the low-energy electron excitations occupy disconnected segments located at the nodal region. In particular, the tips of these disconnected segments converge on the hot spots to form the Fermi pockets, therefore there is a coexistence of the Fermi arcs and Fermi pockets. Moreover, our results indicate that in corresponding to the momentum dependence of the pseudogap, the pseudogap-induced peak-dip-hump (PDH) structure in the electron spectrum is particularly obvious around the antinodal region \cite{Damascelli03,Campuzano04,Carbotte11,Hashimoto15,Matt15,Campuzano99,Fedorov99,Saini97,Ding96,Dessau91,Sakai13}. However, although a weak PDH structure emerges around the nodal region, our theory also predicts that this PDH structure is absent from the hot-spot directions.

The paper is organized as follows. The general formalism of the electron spectral function of the $t$-$J$ model in the fermion-spin representation obtained in terms of the full charge-spin recombination scheme is presented in Sec. \ref{framework}. Within this basic formalism of the electron spectral function, we therefore discuss the nature of the coexistence of the Fermi arcs and Fermi pockets in the pseudogap phase of cuprate superconductors in Sec. \ref{pockets}, where we also show that the single-particle coherent weight is closely related to the pseudogap, and increases with the increase of the doping concentration. In particular, the calculated overall dispersion of the electron excitations is in qualitative agreement with the ARPES experimental data. Finally, we give a summary and discussions in Sec. \ref{conclusions}.

\section{General formalism}\label{framework}

\begin{figure*}[t!]
\centering
\includegraphics[scale=0.70]{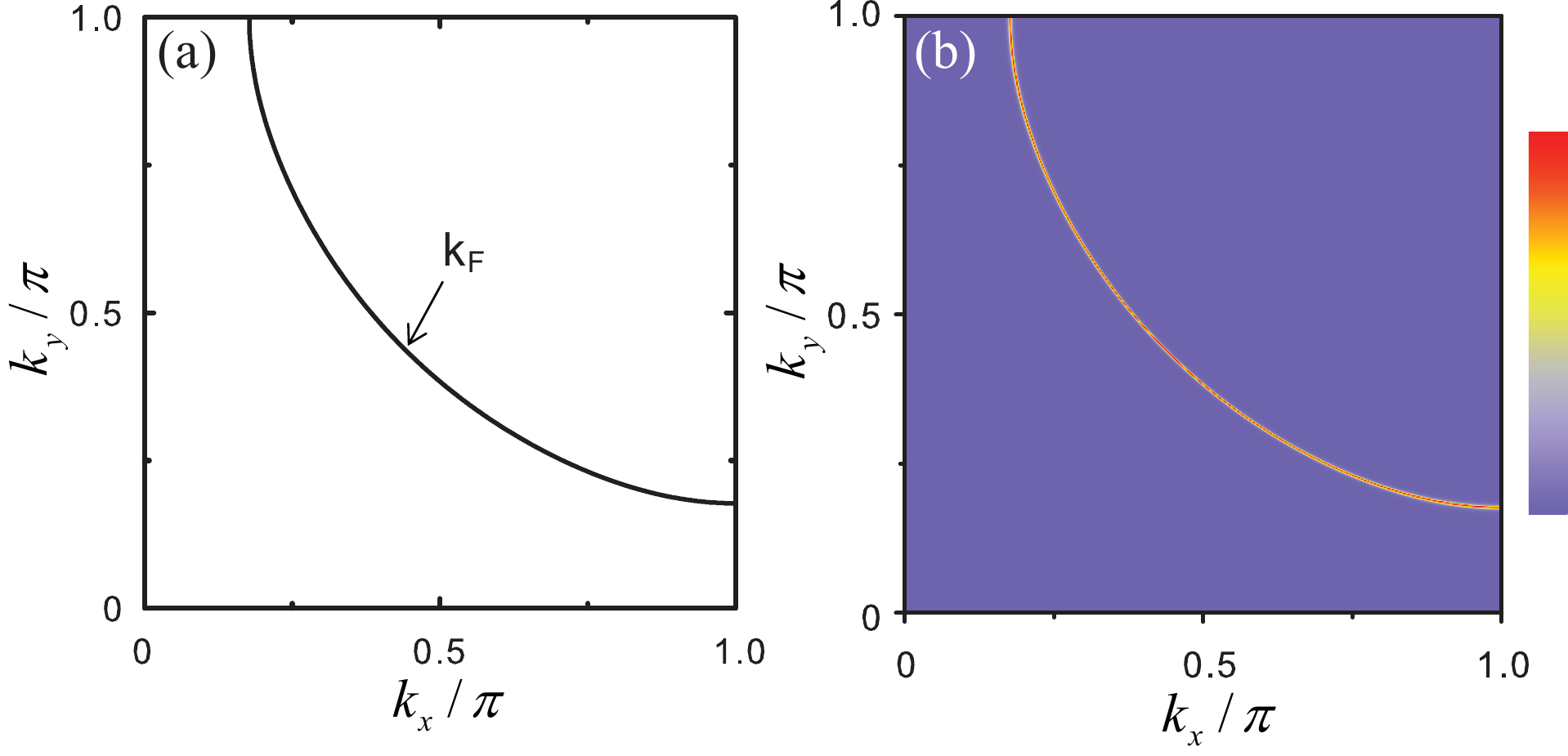}
\caption{(Color online) (a) The contour ${\bf k}_{\rm F}$ and (b) the map of the electron spectral intensity $A({\bf k},0)$ obtained within the mean-field level at $\delta=0.15$ with $T=0.002J$ for $t/J=2.5$ and $t'/t=0.3$.
\label{spectral-maps-MF}}
\end{figure*}

In the parent compound of cuprate superconductors, the hopping of the electrons from site to site is prohibited since the on-site Coulomb repulsive interaction between electrons is much larger than the kinetic energy gain, leading to that the parent compound of cuprate superconductors is a Mott insulator with an antiferromagnetic (AF) long-range order (AFLRO) \cite{Kastner98,Fujita12,Lee88}. However, when this parent Mott insulator is doped with a small percentage of charge carriers, AFLRO is rapidly suppressed leaving the AF short-range order (AFSRO) correlation still intact \cite{Kastner98,Fujita12,Lee88}. In this metallic state with AFSRO, the electrons become mobile but the strong electron correlation from the parent Mott insulating state is thought to survive \cite{Lee88,Yu92,Lee06}, leading to that some competing orders \cite{Comin15,Campi15,Comin14,Wu11,Chang12,Ghiringhelli12,Neto14,Comin15a,Hashimoto15}, pseudogap \cite{Hufner08,Timusk99}, and eventually superconductivity appear. Very soon after the discovery of superconductivity in cuprate superconductors \cite{Bednorz86}, Anderson \cite{Anderson87} argued that this essential physics is contained in the square-lattice $t$-$J$ model acting on the space with no doubly occupied sites $\sum_{\sigma}C^{\dagger}_{l\sigma} C_{l\sigma}\leq 1$, where the electron operators $C^{\dagger}_{l\sigma}$ and $C_{l\sigma}$ that respectively create and annihilate electrons with spin $\sigma$. The strong electron correlation in the $t$-$J$ model manifests itself by this no-double electron occupancy local constraint, and therefore the crucial requirement is to impose this local constraint \cite{Yu92,Lee06,Feng93}. It has been shown that this no-double electron occupancy local constraint can be treated properly in actual calculations within the framework of the charge-spin separation fermion-spin theory \cite{Feng9404,Feng15}, where
the constrained electron is decoupled as a charge carrier and a localized spin, with the charge carrier that represents the charge degree of freedom together with some effects of spin configuration rearrangements due to the presence of the doped charge carrier itself, while the localized spin represents the spin degree of freedom. However, a long-standing unsolved problem is how a microscopic theory based on the charge-spin separation can give a consistent description of the nature of EFS in cuprate superconductors \cite{Yu92,Lee06,Feng93}. To solve this problem, we \cite{Feng15a,Feng16} have developed a full charge-spin recombination scheme to fully recombine a charge carrier and a localized spin into a constrained electron, where the obtained electron propagator can produce a large EFS satisfying Luttinger's theorem. Following our previous discussions \cite{Feng15a,Feng16}, the normal-state electron Green's function of the $t$-$J$ model in the fermion-spin representation can be obtained in terms of the full charge-spin recombination scheme as,
\begin{eqnarray}\label{EGF}
G({\bf k},\omega)={1\over \omega-\varepsilon_{\bf k}-\Sigma_{1}({\bf k},\omega)},
\end{eqnarray}
where $\varepsilon_{\bf k}=-Zt\gamma_{\bf k}+Zt'\gamma_{\bf k}'+\mu$ is the mean-field (MF) electron excitation spectrum, with the chemical potential $\mu$, the nearest-neighbor (NN) and next NN hopping integrals $t$ and $t'$, respectively, $\gamma_{\bf k}=({\rm cos} k_{x}+{\rm cos}k_{y})/2$, $\gamma_{\bf k}'= {\rm cos}k_{x}{\rm cos}k_{y}$, and the number of the NN or next NN sites on a square lattice $Z$. In the framework of the full charge-spin recombination \cite{Feng15a}, the electron self-energy $\Sigma_{1}({\bf k},\omega)$ due to the interaction between electrons by the exchange of spin excitations has been calculated in terms of the spin bubble as \cite{Feng16},
\begin{eqnarray}\label{ESE1}
\Sigma_{1}({\bf k},i\omega_{n})&=&{1\over N^{2}}\sum_{{\bf p,p'}}\Lambda^{2}_{{\bf p}+{\bf p}'+{\bf k}}\nonumber\\
&\times& {1\over \beta}\sum_{ip_{m}}G({\bf p}+{\bf k},ip_{m}+i\omega_{n})\Pi({\bf p}, {\bf p}',ip_{m}),~~~~~
\end{eqnarray}
with $\Lambda_{{\bf k}}=Zt\gamma_{\bf k}-Zt'\gamma_{\bf k}'$, and the spin bubble,
\begin{eqnarray}\label{SB}
\Pi({\bf p},{\bf p}',ip_{m})&=&{1\over\beta}\sum_{ip'_{m}}D^{(0)}({\bf p'},ip_{m}')\nonumber\\
&\times& D^{(0)}({\bf p}'+{\bf p},ip_{m}'+ip_{m}),
\end{eqnarray}
where $D^{(0)-1}({\bf k},\omega)=(\omega^{2}-\omega^{2}_{\bf k})/B_{\bf k}$ is the MF spin Green's function with the MF spin excitation spectrum $\omega_{\bf k}$ and function $B_{\bf k}$ that have been given explicitly in Ref. \cite{Feng15}. In particular, this electron self-energy $\Sigma_{1}({\bf k},\omega)$ has been evaluated explicitly as \cite{Feng16},
\begin{eqnarray}\label{ESE}
\Sigma_{1}({\bf k},\omega)&=&{1\over N^{2}}\sum_{{\bf pp'}\mu\nu}(-1)^{\nu+1}\Omega_{\bf pp'k}\nonumber\\
&\times& {F^{(\nu)}_{{\rm n\mu} {\bf pp'k}}\over \omega+(-1)^{\mu+1} \omega_{\nu{\bf p}{\bf p}'}- \bar{\varepsilon}_{{\bf p}+{\bf k}}},~~~
\end{eqnarray}
where $\mu (\nu)=1,2$, $\Omega_{\bf pp'k}=Z_{\rm F}\Lambda^{2}_{{\bf p}+{\bf p}'+{\bf k}}B_{{\bf p}'} B_{{\bf p}+{\bf p}'}/(4\omega_{{\bf p}'}\omega_{{\bf p}+{\bf p}'})$, $\omega_{\nu{\bf p}{\bf p}'}=\omega_{{\bf p}+{\bf p}'}-(-1)^{\nu}\omega_{\bf p'}$, and $\bar{\varepsilon}_{\bf k}=Z_{\rm F}\varepsilon_{\bf k}$ with the single-particle coherent weight $Z_{\rm F}$ that has been given in Ref. \onlinecite{Feng16}, while the function,
\begin{eqnarray}
F^{(\nu)}_{{\rm n\mu}{\bf pp'k}}=n_{\rm F}[(-1)^{\mu+1}\bar{\varepsilon}_{{\bf p}+{\bf k}}]n^{(\nu)}_{{\rm 1B} {\bf pp'}}+n^{(\nu)}_{{\rm 2B}{\bf pp'}},
\end{eqnarray}
with $n^{(\nu)}_{{\rm 1B}{\bf pp'}}=1+n_{\rm B}(\omega_{{\bf p}'+{\bf p}})+n_{\rm B}[(-1)^{\nu+1}\omega_{\bf p'}]$, $n^{(\nu)}_{{\rm 2B}{\bf pp'}}=n_{\rm B}(\omega_{{\bf p}'+{\bf p}}) n_{\rm B}[(-1)^{\nu+1} \omega_{\bf p'}]$, and $n_{\rm B}(\omega)$ and $n_{\rm F}(\omega)$ that are the boson and fermion distribution functions, respectively. Thus not only the structure of the one-electron bands, but also the structure of the basic interaction are contained in the electron Green's function (\ref{EGF}). With the help of this electron Green's function (\ref{EGF}), the electron spectral function is therefore obtained as,
\begin{eqnarray}\label{SF}
A({\bf k},\omega)={2|{\rm Im}\Sigma_{1}({\bf k},\omega)|\over [\omega-\varepsilon_{\bf k}-{\rm Re}\Sigma_{1}({\bf k},\omega)]^{2}+[{\rm Im}\Sigma_{1}({\bf k},\omega)]^{2}},~~~
\end{eqnarray}
where ${\rm Im}\Sigma_{1}({\bf k},\omega)$ and ${\rm Re}\Sigma_{1}({\bf k},\omega)$ are, respectively, the corresponding imaginary and real parts of $\Sigma_{1}({\bf k},\omega)$.

\section{Nature of electron Fermi surface in pseudogap phase}\label{pockets}

The coexistence of the Fermi arcs and Fermi pockets in the pseudogap phase of cuprate superconductors challenges the traditional concept of an EFS as a continuous contour of the gapless quasiparticle excitations in momentum space that separates the occupied and unoccupied states. However, we in this section will show that the coexistence of the Fermi arcs and Fermi pockets is a natural consequence of the pseudogap-induced EFS instability, and therefore there is a closed connection between the coexistence of the Fermi arcs and Fermi pockets and pseudogap.

\subsection{Coexistence of Fermi arcs and Fermi pockets}

\begin{figure*}[t!]
\centering
\includegraphics[scale=0.80]{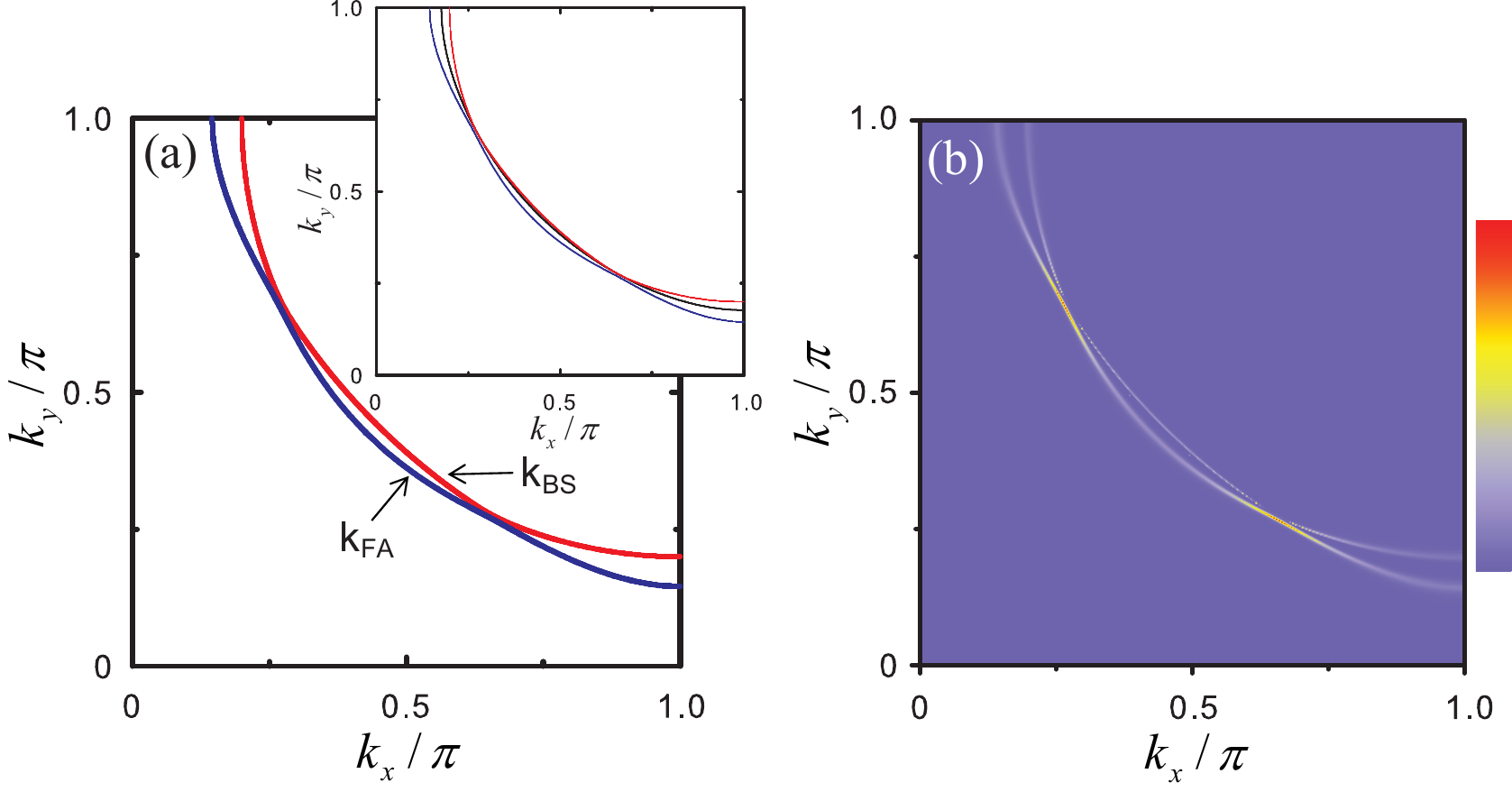}
\caption{(Color online) (a) The contour ${\bf k}_{\rm FA}$ (blue line) and contour ${\bf k}_{\rm BS}$ (red line) obtained from the self-consistent solution (\ref{FS}) and (b) the map of the electron spectral intensity $A({\bf k},0)$ at $\delta=0.15$ with $T=0.002J$ for $t/J=2.5$ and $t'/t=0.3$. Inset in (a): a position comparison for the contours ${\bf k}_{\rm FA}$ (blue line), ${\bf k}_{\rm BS}$ (red line), and ${\bf k}_{\rm F}$ (black line). \label{spectral-maps}}
\end{figure*}

The locations of the continuous contours in momentum space are determined directly by the poles of the electron Green's function (\ref{EGF}) at zero energy,
\begin{eqnarray}\label{FS}
\varepsilon_{\bf k}+{\rm Re}\Sigma_{1}({\bf k},0)=0,
\end{eqnarray}
and then the weight of the low-energy electron excitation spectrum $A({\bf k},0)$ in Eq. (\ref{SF}) at the continuous contours is dominated by the inverse of the imaginary part of the electron self-energy $1/|{\rm Im}\Sigma_{1}({\bf k},0)|$. However, in the static-limit approximation for the electron self-energy $\Sigma_{1}({\bf k},\omega)$ (then within the MF level) \cite{Feng16}, the equation (\ref{FS}) is reduced as $\bar{\varepsilon}_{\bf k}=Z_{\rm F}\varepsilon_{\bf k}=0$, and then the electron spectral function in Eq. (\ref{SF}) is reduced as $A({\bf k},\omega)=2\pi Z_{\rm F}\delta(\omega-\bar{\varepsilon}_{\bf k})$. For a convenience in the following discussions, we plot (a) the continuous contour ${\bf k}_{\rm F}$ in momentum space obtained from the $\varepsilon_{\bf k}=0$ and (b) the map of the electron spectral intensity obtained from $A({\bf k}, \omega)=2\pi Z_{\rm F} \delta(\omega-\bar{\varepsilon}_{\bf k})$ at doping $\delta=0.15$ with temperature $T=0.002J$ for parameters $t/J=2.5$ and $t'/t=0.3$ in Fig. \ref{spectral-maps-MF}. In this static-limit approximation for the electron self-energy, only one continuous contour ${\bf k}_{\rm F}$ is found in momentum space as shown in Fig. \ref{spectral-maps-MF}a, and then the weight of the low-energy quasiparticle excitations is located on this continuous contour ${\bf k}_{\rm F}$ to form a large EFS as Fig. \ref{spectral-maps-MF}b, where the quasiparticle excitation spectrum is gapless, and then the quasiparticle lifetime on the continuous EFS contour ${\bf k}_{\rm F}$ is infinitely long \cite{Feng16}. In particular, this EFS with the area contains $1-\delta$ electrons \cite{Feng16}, and therefore fulfills Luttinger's theorem.

\begin{figure}[h!]
\centering
\includegraphics[scale=0.55]{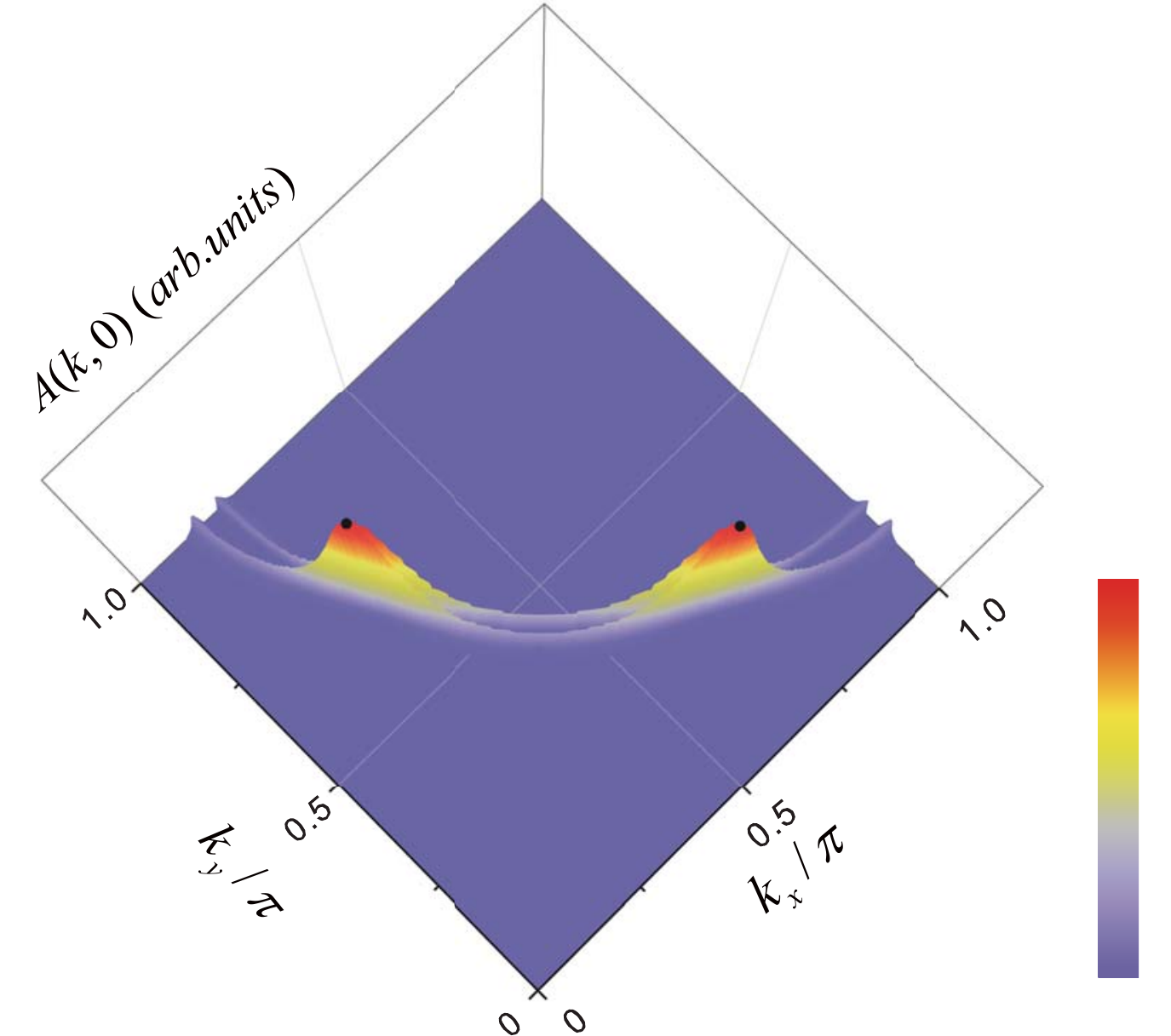}
\caption{(Color online) The electron spectral function $A({\bf k},0)$ in the $[k_{x},k_{y}]$ plane at $\delta=0.15$ with $T=0.002J$ for $t/J=2.5$ and $t'/t=0.3$. The black circles indicate the locations of the hot spots. \label{spectral-maps-3D}}
\end{figure}

However, when the strong electron correlations are included in terms of the electron self-energy $\Sigma_{1}({\bf k},\omega)$, the electron excitation energies are heavily renormalized, which leads to a redistribution of the spectral weight of the low-energy electron excitations. To see this point clearly, we plot (a) the continuous contours in momentum space obtained directly from the self-consistent equation (\ref{FS}) and (b) the map of the electron spectral intensity $A({\bf k},0)$ in Eq. (\ref{SF}) at $\delta=0.15$ with $T=0.002J$ for $t/J=2.5$ and $t'/t=0.3$ in Fig. \ref{spectral-maps}. Obviously, the result of the self-consistent solution in Fig. \ref{spectral-maps}a indicates that there are two continuous contours in momentum space, which are labeled as ${\bf k}_{\rm FA}$ and ${\bf k}_{\rm BS}$, respectively. However, in comparison with the continuous contour ${\bf k}_{\rm F}$ shown in Fig. \ref{spectral-maps-MF}a, we therefore find that both the contours ${\bf k}_{\rm FA}$ and ${\bf k}_{\rm BS}$ are shifted away from the contour ${\bf k}_{\rm F}$ except for the hot spots as shown in the inset of Fig. \ref{spectral-maps}a, since both the contours ${\bf k}_{\rm FA}$ and ${\bf k}_{\rm BS}$ converge on the hot spots. This reflects a fact that the EFS in Fig. \ref{spectral-maps-MF}a in the MF level has been reconstructed by the strong electron correlations. On the other hand, the result in Fig. \ref{spectral-maps}b shows that the low-energy spectral weight at the contours ${\bf k}_{\rm FA}$ and ${\bf k}_{\rm BS}$ around the antinodal region has been suppressed by $|{\rm Im} \Sigma_{1}({\bf k},0)|$, and then the low-energy electron excitations occupy disconnected segments located at the contours ${\bf k}_{\rm FA}$ and ${\bf k} _{\rm BS}$ around the nodal region. To show the spectral weight redistribution on these disconnected segments more clearly, we plot $A({\bf k}, 0)$ in the $[k_{x},k_{y}]$ plane at $\delta=0.15$ with $T=0.002J$ for $t/J=2.5$ and $t'/t=0.3$ in Fig. \ref{spectral-maps-3D}, where the locations of the hot spots, the highest peak heights on the contours, are marked by the black circles. The positions of the hot spots obtained in the present framework are qualitatively consistent with the ARPES experimental data \cite{Sassa11,Comin14}, since these ARPES experimental observations indicate that the sharp quasiparticle peaks with large spectral weights (then the hot spots) appear always at the off-node places. It is thus shown that the tips of the disconnected segments on the contours ${\bf k}_{\rm FA}$ and ${\bf k}_{\rm BS}$ converge at the hot spots to form a closed Fermi pocket, leading to a coexistence of the Fermi arc and Fermi pocket, where the disconnected segment at the first contour ${\bf k}_{\rm FA}$ is so-called the Fermi arc, and is also defined as the front side of the Fermi pocket, while the other at the second contour ${\bf k}_{\rm BS}$ is associated with the back side of the Fermi pocket. However, the result in Fig. \ref{spectral-maps-3D} also shows that the partial spectral-weight at the Fermi arc is transferred to the back side of the Fermi pocket due to the spectral weight redistribution in the presence of the strong electron correlations, which leads to that although both the Fermi arc and back side of the Fermi pocket possess finite spectral weight, the electron excitation peaks at both the Fermi arc and back side of the Fermi pocket are anomalously broad, in qualitative agreement with the ARPES experimental results \cite{Meng09}. On the other hand, at the hot spots, the spectral weight is much larger than these on the Fermi arc and back side of the Fermi pocket, and then the quasiparticle peaks are extremely sharp, this is why these hot spots connected by the charge-order wave vector contribute effectively to the quasiparticle scattering process \cite{Comin15,Campi15,Comin14,Wu11,Chang12,Ghiringhelli12,Neto14,Comin15a,Hashimoto15}. Furthermore, we have also made a series of calculations for $A({\bf k},0)$ at different doping levels, and the results show that the area of the Fermi pockets are proportional to the doping concentration, i.e., in contrast to the case of the doping dependence of the charge-order wave vector $Q_{\rm HS}$ \cite{Feng16}, the Fermi-arc length and area of the Fermi pockets smoothly increases with the increase of doping, and then the Fermi arc length evolves into a continuous EFS contour in momentum space in the heavily overdoped regime.

\begin{figure*}[t!]
\centering
\includegraphics[scale=0.75]{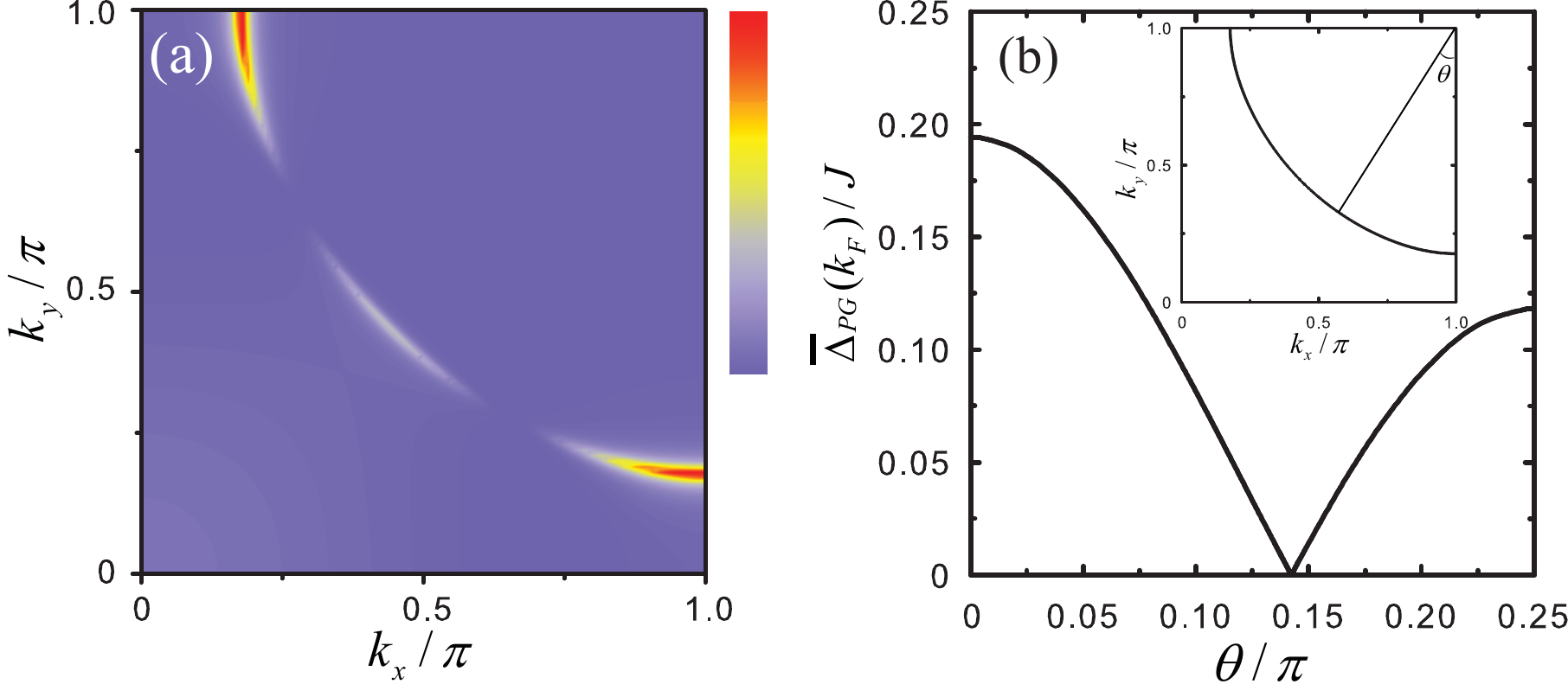}
\caption{(Color online) (a) The map of the imaginary part of the electron self-energy (b) the angular dependence of the pseudogap on the electron Fermi surface at $\delta=0.15$ with $T=0.002J$ for $t/J=2.5$ and $t'/t=0.3$. \label{imaginary-part-self-energy-PG}}
\end{figure*}

The essential physics of the coexistence of the Fermi arcs and Fermi pockets is closely related to the emergence of the momentum dependence of the pseudogap. This follows a fact that the electron self-energy $\Sigma_{1}({\bf k},\omega)$ in Eq. (\ref{ESE}) can be also rewritten as \cite{Feng16,Feng15a},
\begin{eqnarray}\label{PG}
\Sigma_{1}({\bf k},\omega)\approx {[\bar{\Delta}_{\rm PG}({\bf k})]^{2}\over\omega+\varepsilon_{0{\bf k}}},
\end{eqnarray}
with the energy spectrum $\varepsilon_{0{\bf k}}=L^{({\rm e})}_{2}({\bf k})/L^{({\rm e})}_{1}({\bf k})$, the pseudogap $\bar{\Delta}_{\rm PG}({\bf k})=L^{({\rm e})}_{2}({\bf k})/ \sqrt{L^{({\rm e})}_{1}({\bf k})}$, the functions $L^{({\rm e})}_{1}({\bf k})=-\Sigma_{\rm 1o}({\bf k},\omega=0)$ and $L^{({\rm e} )}_{2}({\bf k})=\Sigma_{1}({\bf k},\omega=0)$, while $\Sigma_{\rm 1o}({\bf k},\omega=0)$ is the antisymmetric part of the electron self-energy, and it and the functions $L^{({\rm e})}_{1} ({\bf k})$ and $L^{({\rm e})}_{2}({\bf k} )$ can be obtained directly from $\Sigma_{1}({\bf k},\omega)$ in Eq. (\ref{ESE}). As we \cite{Feng16} have shown in the previous discussions, this pseudogap $\bar{\Delta}_{\rm PG}({\bf k})$ is therefore identified as being a region of the electron self-energy effect in which the pseudogap $\bar{\Delta}_{\rm PG}({\bf k})$ suppresses the spectral weight. It should be emphasized that the equation (\ref{PG}) is exact mapping when $\omega=0$, however, it is a very good approximation for the qualitative description of the low-energy electronic state behavior of cuprate superconductors in the pseudogap phase. In this case, the imaginary part of $\Sigma_{1}({\bf k},\omega)$ can be expressed in terms of the pseudogap as,
\begin{eqnarray}\label{IESE}
{\rm Im}\Sigma_{1}({\bf k},\omega) \approx  2\pi[\bar{\Delta}_{\rm PG}({\bf k})]^{2}\delta(\omega+\varepsilon_{0{\bf k}}),
\end{eqnarray}
which is in agreement with the ARPES experiment \cite{Matt15}, where an intrinsic relation between the electron scattering and pseudogap has been observed. Substituting the electron self-energy $\Sigma_{1}({\bf k},\omega)$ in Eq. (\ref{PG}) into Eq. (\ref{EGF}), the electron Green's function in Eq. (\ref{EGF}) can be rewritten as,
\begin{eqnarray}\label{EGF1}
G({\bf k},\omega)={W^{+}_{\bf k}\over\omega-E^{+}_{\bf k}}+{W^{-}_{\bf k}\over\omega -E^{-}_{\bf k}},
\end{eqnarray}
where $W^{+}_{\bf k}=(E^{+}_{\bf k}+\varepsilon_{0{\bf k}})/(E^{+}_{\bf k}-E^{-}_{\bf k})$ and $W^{-}_{\bf k}=-(E^{-}_{\bf k}+\varepsilon_{0{\bf k}})/(E^{+}_{\bf k}-E^{-}_{\bf k})$ are the coherence factors, and satisfy the sum rule: $W^{+}_{\bf k}+W^{-}_{\bf k}=1$. As a consequence of the presence of the pseudogap, the electron energy band has been split into the antibonding band $E^{+}_{\bf k}=[\varepsilon_{\bf k}-\varepsilon_{0{\bf k}}+\sqrt{(\varepsilon_{\bf k}+\varepsilon_{0{\bf k}})^{2}+4\bar{\Delta}^{2} _{\rm PG}({\bf k})}]/2$ and bonding band $E^{-}_{\bf k}=[\varepsilon_{\bf k}-\varepsilon_{0{\bf k}}-\sqrt{(\varepsilon_{\bf k}+\varepsilon_{0{\bf k}})^{2}+4\bar{\Delta}^{2}_{\rm PG}({\bf k})} ]/2$, respectively.

With the above expression of the electron Green's function in Eq. (\ref{EGF1}), now we find that the first contour ${\bf k}_{\rm FA}$, shown as a blue line curve in Fig. \ref{spectral-maps}a, represents the contour in momentum space, where the electron antibonding dispersion $E^{+}_{\bf k}$ along ${\bf k}_{\rm FA}$ is equal to zero, while the second contour ${\bf k}_{\rm BS}$, shown as a red line curve in Fig. \ref{spectral-maps}a, is the contour in momentum space, where the electron bonding dispersion $E^{-}_{\bf k}$ along ${\bf k}_{\rm BS}$ is equal to zero. Since the electron self-energy $\Sigma_{1}({\bf k},\omega)$ originates in the electron's coupling to spin excitations, the pseudogap $\bar{\Delta}_{\rm PG}({\bf k})$ is strong dependence of momentum. To see this strongly anisotropic pseudogap in momentum space clearly, we plot (a) the map of the intensity of $|{\rm Im} \Sigma_{1}({\bf k},0)|$ and (b) the angular dependence of the pseudogap $\bar{\Delta}_{\rm PG}({\bf k}_{\rm F})$ on EFS at $\delta=0.15$ with $T=0.002J$ for $t/J=2.5$ and $t'/t=0.3$ in Fig. \ref{imaginary-part-self-energy-PG}. It is shown clearly that as in the case of $|{\rm Im}\Sigma_{1}({\bf k},0)|$, the pseudogap $\bar{\Delta}_{\rm PG}({\bf k}_{\rm F})$ has a strong angular dependence with the actual maximum at the antinode, the Fermi momentum on the BZ boundary, which leads to that the low-energy electron spectral weight at the contours ${\bf k}_{\rm FA}$ and ${\bf k}_{\rm BS}$ around the antinodal region is gapped out by the pseudogap. However, the actual minimum does not appear around the node, but locates exactly at the hot spot ${\bf k}_{\rm HS}$, where the pseudogap $\bar{\Delta}_{\rm PG}({\bf k}_{\rm HS})=0$, the energy spectra $\varepsilon_{0{\bf k}_{\rm HS}} =-\varepsilon_{{\bf k}_{\rm HS}}$, and then $E^{+}_{{\bf k}_{\rm HS}}=E^{-}_{{\bf k}_{\rm HS}}=\varepsilon_{{\bf k}_{\rm HS}}$ for the electron quasiparticle excitation spectra at the antibonding and bonding bands, which leads to that the tips of these disconnected segments on ${\bf k}_{\rm FA}$ and ${\bf k}_{\rm BS}$ converge on the hot spots to form the closed Fermi pocket around the nodal region, generating a coexistence of the Fermi arcs and Fermi pockets. In other words, the coexistence of the Fermi arcs and Fermi pockets is a natural consequence of the emergence of the strong momentum dependence of the pseudogap. Furthermore, the magnitude of the pseudogap parameter $\bar{\Delta}_{\rm PG}$ is the doping dependent, and smoothly decreases upon increasing doping \cite{Feng16,Feng15a}. This doping dependence of the pseudogap $\bar{\Delta}_{\rm PG}$ therefore leads to that the Fermi-arc length (then the area of the Fermi pockets) increases with the increase of doping, and then it covers the full length of EFS in the heavily overdoped regime.

\begin{figure*}[t!]
\centering
\includegraphics[scale=0.75]{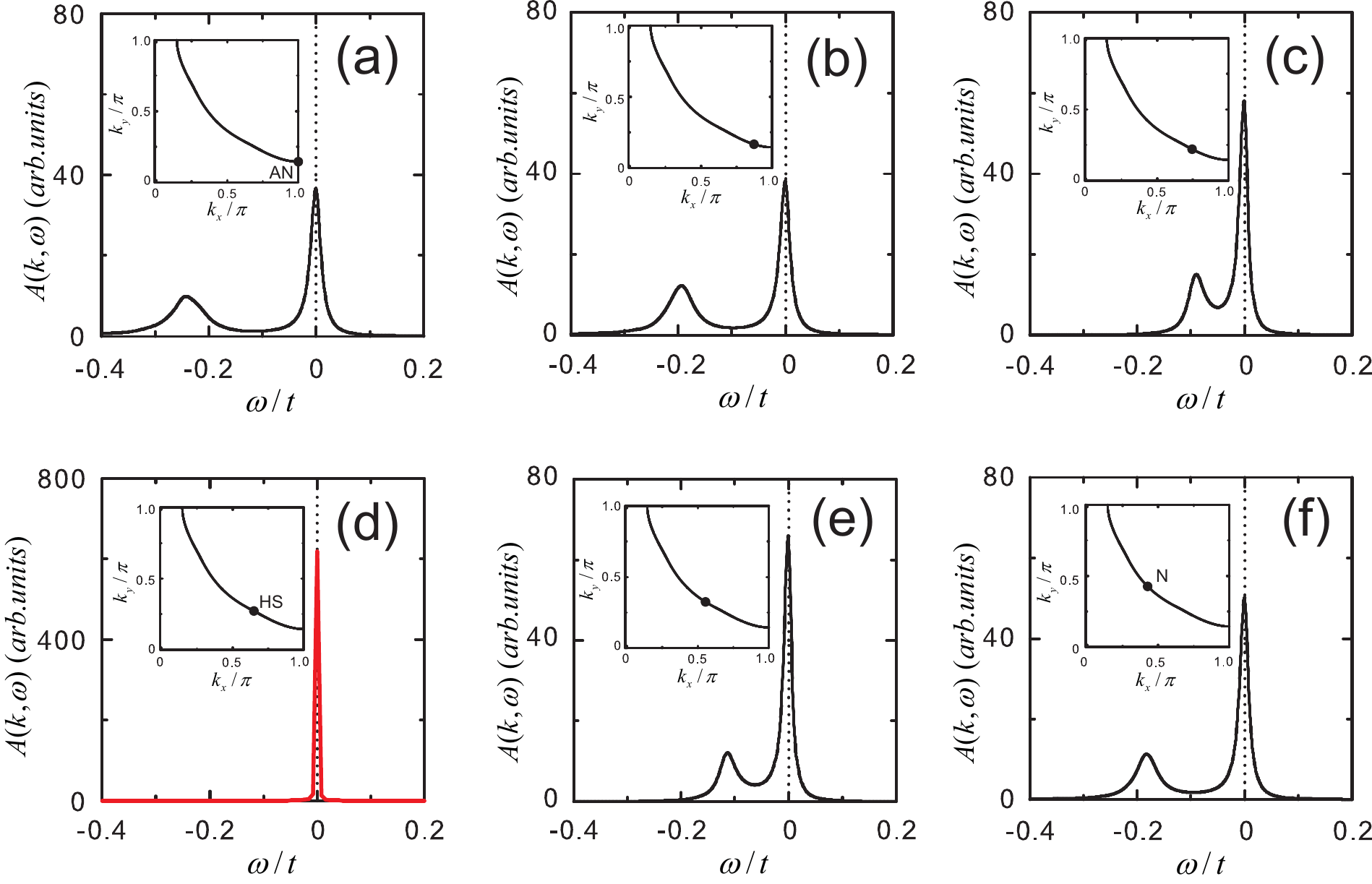}
\caption{(Color online) The electron spectral function at the contour ${\bf k}_{\rm FA}$ in $\delta=0.15$ with $T=0.002J$ for $t/J=2.5$ and $t'/t=0.3$, taken at the point cut in the Brillouin zone inset to the figures, where AN, HS, and N denote the antinode, hot spot, and node, respectively. \label{PDH}}
\end{figure*}

\subsection{Peak-dip-hump structure in electron spectrum}

In the early days of the electronic structure of cuprate superconductors, the ARPES experiments observed a tremendous change in the electron excitation spectral lineshape around the antinodal region, where an electron excitation peak develops at the lowest binding energy, followed by a dip and a hump, giving rise to the remarkable PDH structure in the electron excitation spectrum \cite{Damascelli03,Campuzano04,Carbotte11,Hashimoto15,Matt15,Campuzano99,Fedorov99,Saini97,Ding96,Dessau91,Sakai13}. Later, the ARPES measurements have been extended to study the doping, temperature, and momentum dependence of the electron excitation spectrum and found that (a) the hump scales with the peak and persists above $T_{\rm c}$ in the pseudogap phase \cite{Campuzano99}, reflecting that the well pronounced PDH structure is totally unrelated to superconductivity; (b) although the PDH structure is most strongly developed around the antinodal region, the weak PDH structure was also observed around the nodal region \cite{Sakai13}; (c) the PDH structure is mainly caused by the pseudogap \cite{Hashimoto15,Matt15}. The study of the electron excitation spectrum of cuprate superconductors thus is complicated due to the presence of the strong doping, temperature, and momentum dependence of the pseudogap. As a complement of the above analysis of the pseudogap-generated coexistence of the Fermi arcs and Fermi pockets, we in this subsection study the nature of the PDH structure of the electron excitation spectrum in the normal-state pseudogap phase, and show that as a natural consequence of the momentum dependence of the pseudogap shown in Fig. \ref{imaginary-part-self-energy-PG}b, the PDH structure is thus absent from the hot spot directions. To show this absence of the PDH structure from the hot spot directions clearly, we have made a series of calculations for the electron excitation spectral function $A({\bf k},\omega)$ at the contour ${\bf k}_{\rm FA}$ along the momentum from the antinode to node, and the results are plotted in Fig. \ref{PDH}. It is obvious that at the antinode, an additional peak in the electron excitation spectrum appears at the higher energy region (see Fig. \ref{PDH}a), however, the low-energy peak is much sharper than this additional peak. In this case, the electron excitation spectrum consists of two peaks, with a low-energy peak, and a weak high-energy peak, which is associated with the hump, while the spectral dip is in between them, and then the total contributions for the electron excitation spectrum give rise to the PDH structure, in good agreement with the ARPES experimental observations on cuprate superconductors \cite{Damascelli03,Campuzano04,Carbotte11,Hashimoto15,Matt15,Campuzano99,Fedorov99,Saini97,Ding96,Dessau91,Sakai13}. However, the position of this weak high-energy hump is momentum dependent, i.e., when the momentum moves away from the antinode and towards to the hot spot, the position of the hump appreciably shift to the low-energy peak (see Fig. \ref{PDH}b and Fig. \ref{PDH}c). In particular, this hump is incorporated with the low-energy peak at the hot spot (see Fig. \ref{PDH}d), leading to an absence of the PDH structure at the hot spot directions. However, this PDH structure develops again when the momentum moves away from the hot spot and towards to the node (see Fig. \ref{PDH}e), and then a weak PDH structure appears at the nodal point (see Fig. \ref{PDH}f). As shown in Fig. \ref{imaginary-part-self-energy-PG}b, the magnitude of the pseudogap exhibits the largest value at the antinode, and then it decreases with the move of the momentum from the antinode to the hot spot. On the other hand, the magnitude of the pseudogap reaches its minimum at the hot spot, and then it increases with the move of the momentum from the hot spot to the node. This special momentum dependence of the pseudogap therefore induces the striking feature of the PDH structure in the electron excitation spectrum, i.e., the PDH structure is particularly obvious around the antinodal region, then it disappears at the hot spot directions, and eventually a weak PDH structure emerges around the nodal region.

\subsection{Dispersion of electron excitations}

\begin{figure*}[t!]
\centering
\includegraphics[scale=0.65]{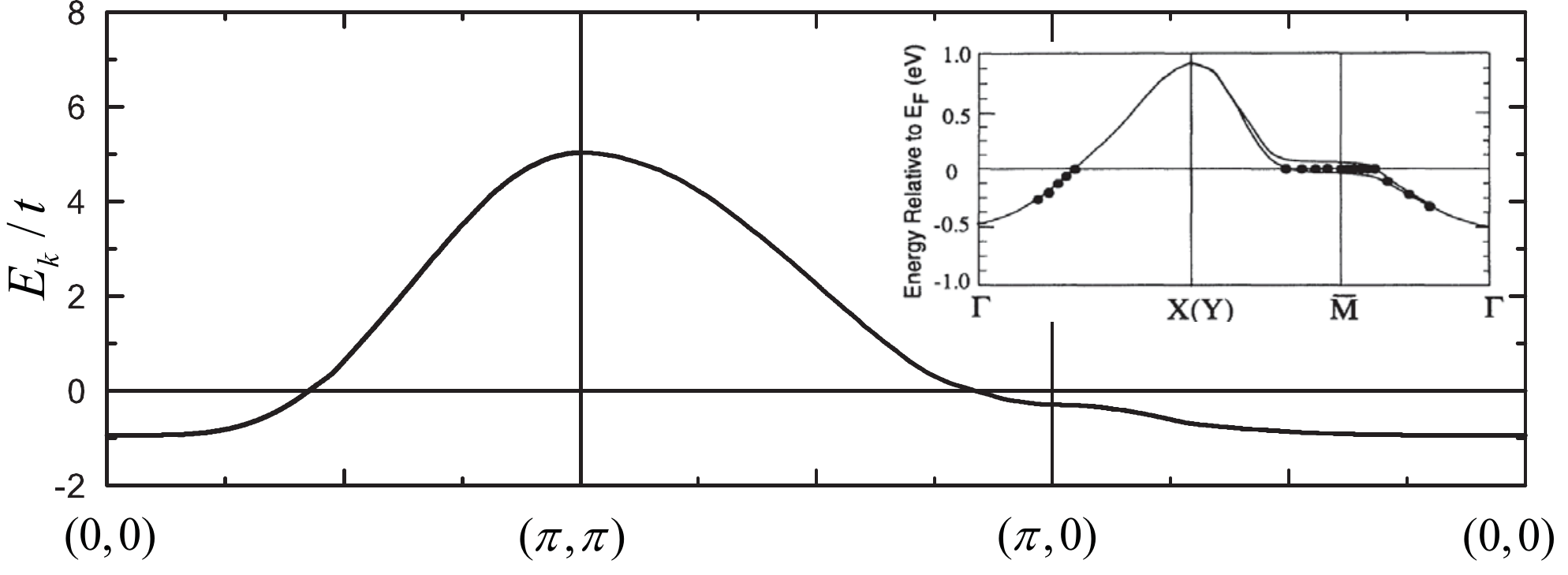}
\caption{The position of the lowest-energy solutions of the self-consistent equation (\ref{band}) as a function of momentum at $\delta=0.15$ with $T=0.002J$ for $t/J=2.5$ and $t'/t=0.3$. Inset: the corresponding experimental results of Bi$_{2}$Sr$_{2}$CaCu$_{2}$O$_{8+\delta}$ taken from Ref. \onlinecite{Dessau93}, where the solid circles represent the experimental data points, while the solid lines represent the best interpretation of the dispersion relationships. \label{dispersion}}
\end{figure*}

The poles of the electron Green's function (\ref{EGF}) map the energy versus momentum dependence of the electron excitations, i.e., the electron excitation energies are obtained by the solution of the self-consistent equation,
\begin{eqnarray}\label{band}
E_{\bf k}-\varepsilon_{\bf k}-{\rm Re}\Sigma_{1}({\bf k},E_{\bf k})=0,
\end{eqnarray}
and then these energies can be measured in ARPES experiments \cite{Damascelli03,Campuzano04,Kim98,Dessau93,Wells95,Armitage01}. For a further understanding of the unusual feature of the electron excitations of cuprate superconductors in the normal-state pseudogap phase, we plot the positions of the lowest-energy solutions in Eq. (\ref{band}) as a function of momentum along the high symmetry directions of BZ at $\delta=0.15$ with $T=0.002J$ for $t/J=2.5$ and $t'/t=0.3$ in Fig. \ref{dispersion} in comparison with the corresponding experimental data \cite{Dessau93} of Bi$_{2}$Sr$_{2}$CaCu$_{2}$O$_{8+\delta}$ (inset). Apparently, our present theoretical result captures the qualitative feature of the overall dispersion of the electron excitations observed experimentally on cuprate superconductors in the normal-state pseudogap phase \cite{Damascelli03,Campuzano04,Kim98,Dessau93,Wells95,Armitage01}. In corresponding to the large values of the pseudogap around the antinodal region as shown in Fig. \ref{imaginary-part-self-energy-PG}b, the dispersion of the electron excitations around the $[\pi,0]$ point has an anomalously small changes of electron energy as a function of
momentum, leading to appearance of the unusual flat band around the $[\pi,0]$ point. In particular, this flat band is slightly below the Fermi energy. These theoretical results are in qualitative agreement with the experimental results \cite{Damascelli03,Campuzano04,Kim98,Dessau93,Wells95,Armitage01}. Our results also show that this unusual flat band is manifestation of a strong coupling between the electron excitations and collective spin excitations.

\subsection{Doping dependence of single-particle coherence}

\begin{figure}[h!]
\centering
\includegraphics[scale=0.60]{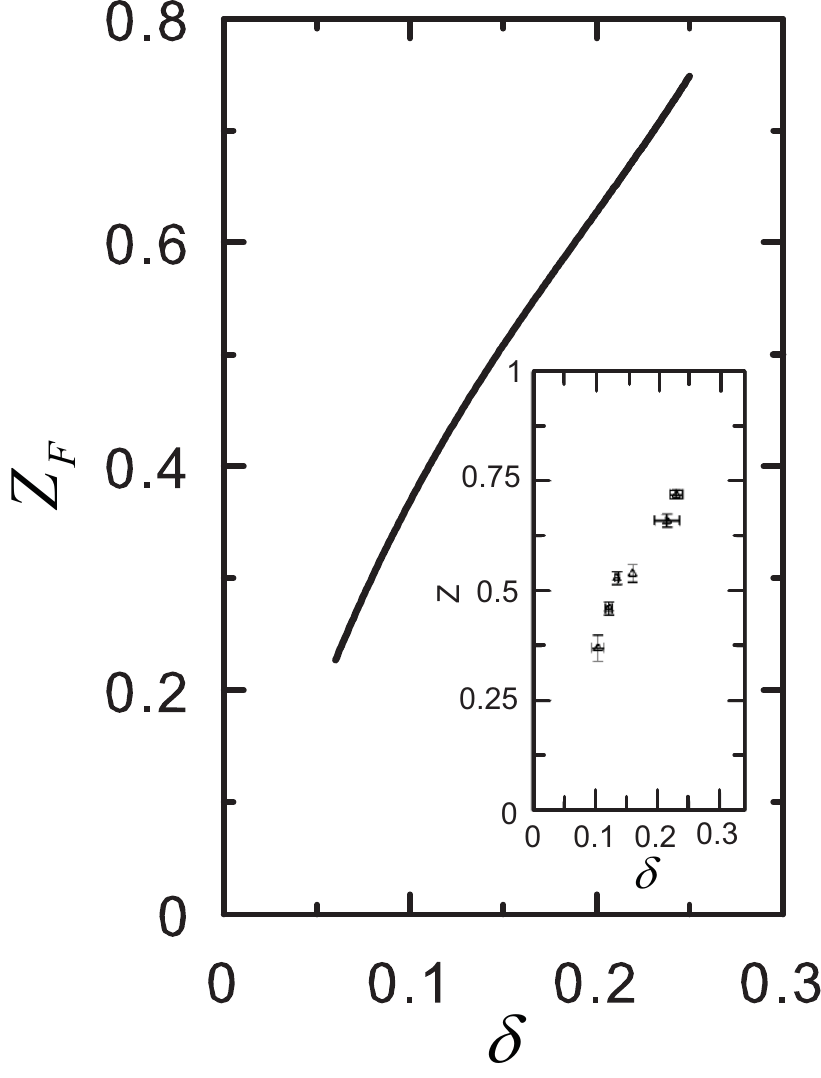}
\caption{The nodal region single-particle coherent weight near the electron Fermi surface as a function of doping with $T=0.002J$ for $t/J=2.5$, $t'/t=0.3$. Inset: the corresponding experimental result of the single-particle coherent weight of Bi$_{2}$Sr$_{2}$CaCu$_{2}$O$_{8+\delta}$ taken from Ref. \onlinecite{Johnson01}. \label{ZFC}}
\end{figure}

Now we turn to discuss the doping dependence of the single-particle coherence of cuprate superconductors in the normal-state pseudogap phase. The doping dependence of the behavior of the single-particle coherence may discriminate whether superconductivity in cuprate superconductors is linked to a doped Mott insulator or not \cite{Fournier10}. The single-particle coherent weight is defined as $Z^{-1}_{\rm F}({\bf k},\omega)=1-{\rm Re}\Sigma_{\rm 1o}({\bf k},\omega)$, where ${\rm Re}\Sigma_{\rm 1o}({\bf k},\omega)$ is the antisymmetric part of the electron self-energy $\Sigma_{1}({\bf k},\omega)$. In the following discussions, we only focus on the low-energy behavior of the single-particle coherence, and in this case, the single-particle coherent weight can be studied in the static limit, i.e., $Z^{-1}_{\rm F}({\bf k})=1-{\rm Re}\Sigma_{\rm 1o}({\bf k},\omega)|_{\omega=0}$. In this static-limit approximation, the single-particle coherent weight is obtained directly from the electron self-energy in Eq. (\ref{PG}) as,
\begin{eqnarray}\label{ZF}
Z^{-1}_{\rm F}({\bf k})=1+{[\bar{\Delta}_{\rm PG}({\bf k})]^{2}\over\varepsilon^{2}_{0{\bf k}}},
\end{eqnarray}
which shows that the partial pseudogap effects have been contained in $Z_{\rm F}({\bf k})$, and then $Z_{\rm F}$ plays a similar role as the pseudogap, i.e., it reduces the electron energy bandwidth, and suppresses the low-energy spectral weight of the electron excitation spectrum. In Fig. \ref{ZFC}, we plot the nodal region $Z_{\rm F}$ near ${\bf k}_{\rm F}$ as a function of doping with $T=0.002J$ for $t/J=2.5$ and $t'/J=0.3$ in comparison with the corresponding experimental result \cite{Johnson01} of the single-particle coherent weight obtained from Bi$_{2}$Sr$_{2}$CaCu$_{2}$O$_{8+\delta}$ (inset). Our present calculations therefore reproduce qualitatively the experimental result of the single-particle coherent weight of cuprate superconductors \cite{Johnson01,Ding01,DLFeng00}. In particular, this single-particle coherent weight $Z_{\rm F}$ is vanishingly small in the heavily underdoped regime, and is increased linearly with doping, i.e., $Z_{\rm F} \propto\delta$, which indicates that only $\delta$ number of the coherently doped carriers are recovered in the normal-state, consistent with the picture of a doped Mott insulator with $\delta$ charge carriers \cite{Anderson87} and the experimental data \cite{Johnson01,Ding01,DLFeng00}.

\section{Conclusions}\label{conclusions}

\begin{figure}[h!]
\centering
\includegraphics[scale=0.40]{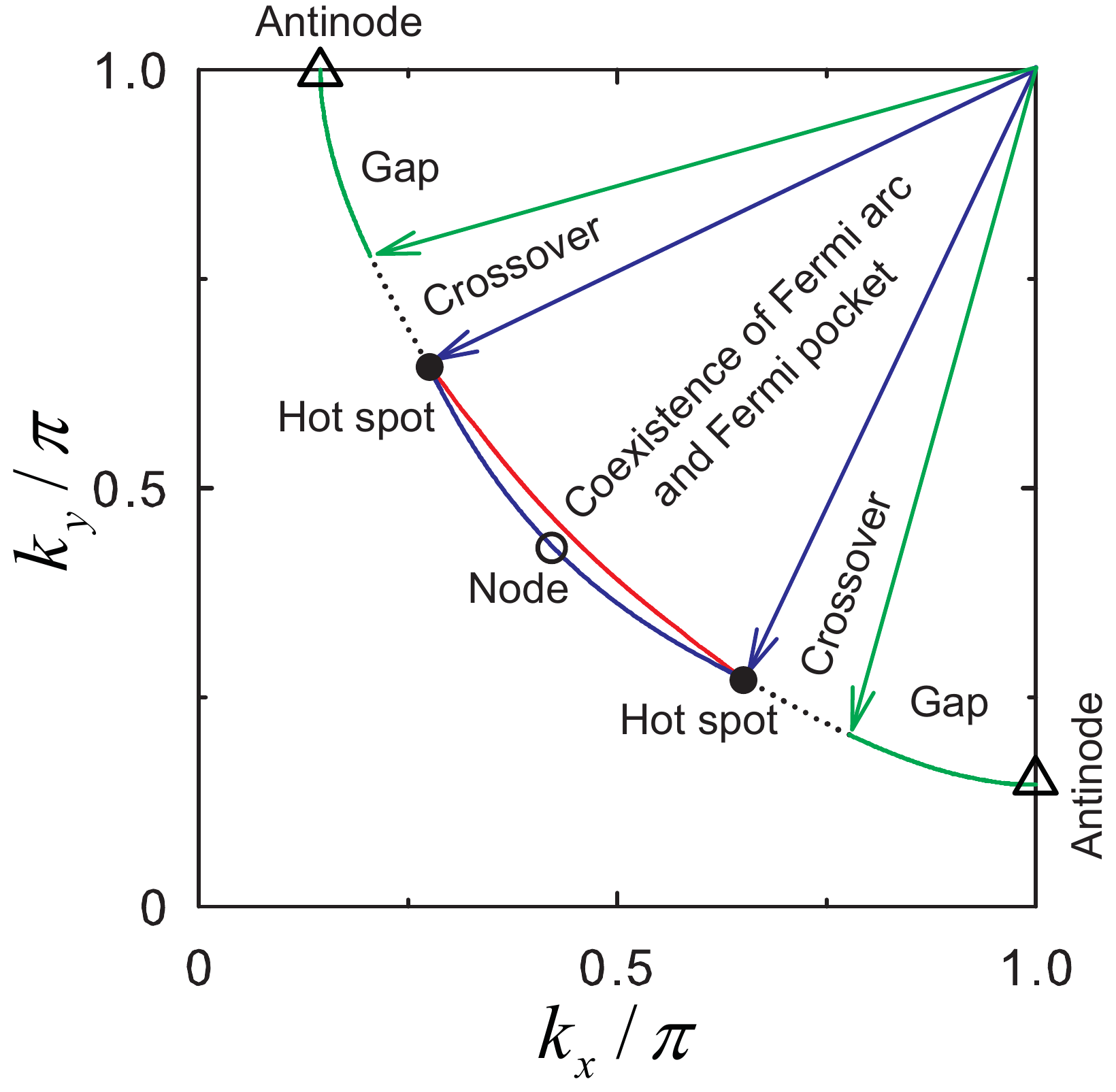}
\caption{(Color online) Schematic electronic structure of cuprate superconductors in the normal-state pseudogap phase. The curve represents the electron Fermi surface in one quadrant of momentum space, defined by momenta in the $[k_{x},k_{y}]$ plane. The hot spots, where the pseudogap vanishes, are marked by the black dots. The Fermi arc and back side of the Fermi pocket around the nodal region are marked in the solid blue and red lines, respectively, where the Fermi pocket coexists with the Fermi arc. The antinodal regions are marked in the solid green lines, where the pseudogap exhibits the largest value at the antinodes. The dashed black line regions mark a crossover between these two regimes. The empty triangles and empty circle denote the antinodes and node, respectively. \label{phase-diagram}}
\end{figure}

In conclusion, based on the $t$-$J$ model in the fermion-spin representation, we have studied the origin of the coexistence of the Fermi arcs and Fermi pockets in the normal-state pseudogap phase of cuprate superconductors by taking into account the pseudogap effect, and the main results are summarized in Fig. \ref{phase-diagram}. Our results show that the coexistence of the Fermi arcs and Fermi pockets is a natural consequence of the pseudogap-induced electron spectral-weight redistribution in the normal-state pseudogap phase, where the pseudogap induces an energy band splitting, and then the poles of the electron Green's function at zero energy form two continuous contours in momentum space, however, the low-energy spectral weight at these two contours around the antinodal region is gapped out by the momentum dependence of the pseudogap, and then the low-energy electron excitations occupy disconnected segments located at the nodal region. In particular, the tips of these disconnected segments converge on the hot spots to form the Fermi pockets, therefore there is a coexistence of the Fermi arcs and Fermi pockets. Furthermore, we have shown that the single-particle coherent weight is directly related to the pseudogap, and grows linearly with doping. Moreover, the calculated result of the overall dispersion of the electron excitations is in qualitative agreement with the experimental data. Although the pseudogap-induced PDH structure in the electron spectrum is most strongly developed around the antinodal region, and remains around the nodal region, our theory predicts that this PDH structure is absent from the hot-spot directions. Our result also shows that the coexistence of the Fermi arcs and Fermi pockets is the part of a rich phenomenology associated with the pseudogap physics.

\acknowledgments

The authors would like to thank Dr. L\"ulin Kuang, Professor Yu Lan, and Professor M. Shi for helpful discussions. We also thank Professor M. Shi for bringing Refs. \onlinecite{Shi08} and \onlinecite{Sassa11} to our attention. HZ is supported by the National Natural Science Foundation of China (NSFC) under Grant No. 11547034, and DG and SF are supported by the National Key Research and Development Program of China under Grant No. 2016YFA0300304, and NSFC under Grant Nos. 11274044 and 11574032.

\end{document}